# Deposition of $La_2Zr_2O_7$ Film by Chemical Solution Deposition

Z.M. Yu[1], P. Odier[2,3,*], S. Morlens[2,3], P. Chaudouët[3,4], M. Bacia[4], L. Zhou[1], P.X. Zhang[1], L.H. Jin[1], C.S. Li[1], P. David[2], O. Fruchart[2], Y.F. Lu[1]

1. Northwest Institute for Nonferrous Metal Research, P.O.Box 51, Xi'an, Shaanxi, 710016, P.R.China

2. Néel Institut, CNRS&UJF, 25 avenue des Martyrs, BP166, 38042 Grenoble Cedex, France.

3. CRETA, CNRS, 25 avenue des Martyrs, BP166, 38042 Grenoble Cedex, France.

4. Laboratoire des Matériaux et du Génie Physique - BP 257- INPGrenoble Minatec - 3 parvis Louis Néel - 38016 Grenoble – France.

* Corresponding author: P. Odier, philippe.odier@grenoble.cnrs.fr, Néel Institut (département MCBT, bât.E)/CRETA (Bât. V), CNRS, 25 av des martyrs, BP 166, F-38042 Grenoble cedex 09, France, tel: (33) (0)4 76 88 90 34, fax: (33) (0)4 76 88 12 80.

## Abstract

$La_2Zr_2O_7$ (LZO) formation of bulk powders and of films by Chemical Solution Deposition (CSD) process have been studied using propionates. The treatment involved a one step cycle in the reducing forming gas (Ar-5%$H_2$) to be compatible with Ni-5at%W RABITS. Large amount of residual carbon was found in LZO powders formed in these conditions (10 wt %). The volume fraction of the cube texture in LZO films on Ni-5at%w RABITS was found to be a function of the speed of the gas flown above sample. This phenomenon is discussed in considering the C deposited from the carbon-containing gases emitted during the pyrolysis of the precursor. Using proper conditions (950 °C and the speed of gas of $6.8 \times 10^{-2}$ m/s), LZO films with good surface crystallinity could be obtained on Ni-5at%W RABITS as demonstrated by X-ray diffraction, electron backscattered diffraction and RHEED. The existence of residual carbon in oxide films is a common question to films deposited by CSD processes under reducing condition.

# 1. Introduction

The second generation of high-temperature superconducting tapes, also named YBa$_2$Cu$_3$O$_7$ (YBCO) coated conductors (CCs), sustains a considerable interest and many challenges in the material science community [1]. YBCO is very anisotropic and needs to be cube-textured along its principal axes ([001] and [100]) to allow high superconducting currents across its grain boundaries [2]. The CCs tapes must be flexible for their assembling into cables used in electricity transportation [3]. Films technologies to deposit YBCO layer on a flexible metallic substrate only permit to fulfil these constrains. Ni rolled assisted bi-axially textured substrates (RABITS) are attractive substrates [4], but several buffer layers between the metallic substrate and the YBCO layer are very important because almost all of metallic elements in the metallic substrate are harmful to the superconductivity of YBCO layer. Quite complex architectures of buffer layers have been successfully used on NiW alloys such as those developed by American Superconductors [4] or others.

LZO is a highly refractory oxide (high melting point over 2200°C) and shows interesting properties in radiation tolerance [5], ionic conductivity [6] in doped compositions, thermal barrier coating [7]. Besides these application fields, LZO is also an important material to be used for buffer layers. The simplest low cost architecture with only one buffer made of La$_2$Zr$_2$O$_7$ (LZO) has been recently validated using an original combination of chemical methods involving metalorganic chemical vapour deposition (MOCVD) and metal organic decomposition (MOD); i.e. YBCO$_{MOCVD}$/LZO$_{MOD}$/NiW $_{RABITS}$ [8].

Cost, scalability and simplicity of process are essential ingredients to be considered for long-length CCs tape production. Chemical solution deposition (CSD) is very attractive in this respect and much used for functional oxide films synthesis [9, 10] in three principal approaches: sol-gel, metal-organic decomposition (MOD) and hybrid routes. Sol-gel uses controlled hydrolysis-condensation reactions to build up a 3D organic-inorganic network [11], MOD consists in pyrolysing a metal-organic precursor after its drying into a dense glaze followed by its crystallisation. MOD is often preferred to prepare epitaxial films because the 3D network has been suspected to be detrimental to epitaxial growth [12]. However MOD needs a fine control of the decomposition step and for its crystallisation into oxide, both transformations which are complex to study.

Several important works have been done to deposit LZO films for buffer layers by CSD approaches [13, 14, 15]. However, the reactions leading to the formation of LZO are not clearly established yet as stress the following, but non exhaustive, list of open questions. *First*, the precursor structure itself is often unknown while it might influence the molecular mixing at a local level with a possible influence on the microstructure. *Second*, the pyrolysis of the precursor must be conducted under Ar-5% $H_2$ to avoid any oxidation of the metallic substrate, then reduction of the evolved molecules can be anticipated with important consequences on the crystallisation of the oxide. C deposit has been observed by Rutherford back scattering [14] in such a process (depending however of the forming atmosphere); C could play an important role in the growth [16] and in the build up of the microstructure as observed in a similar case [17, 18]. C was removed efficiently by adding humidity in the forming gas (Ar-5 % $H_2$) but this destroyed the texture [14] ((222) orientation grains were enhanced and (400) decreased). *Third*, while the texture of LZO films obtained on Ni-5at%W substrates reproduces very well that of the substrates [8, 19, 20], the exact mechanism of epitaxy is still an open question [21] since a high structural mismatch (($\varepsilon_{LZO}-\varepsilon_{Ni})/\varepsilon_{Ni} \sim 8\%$) exists with the substrate. However, it is reasonable to admit that the nucleation starts at the interface with the substrate. This is supported by the fact that, at 800 °C, LZO was observed strongly c-axis textured by X ray diffraction (XRD) but amorphous at its surface according to surface probing by reflexion high energy electron diffraction (RHEED); it became bi-axially textured up to its surface after annealing at 900 °C or above [15]. The crystallisation has started at the metal/oxide interface and has developed at higher temperature by growing within the amorphous volume of the film up to its surface. *Fourth*, the microstructure shows rather peculiar properties. Polygonal nanovoids have been identified by transmission electron microscopy (TEM) [21], suspected but not discussed in Ref. 22, and recently studied in more details by TEM [23]. These nanovoids play an important role since they reduce the effective length of matter protecting the metallic substrate from oxidation during the deposition of YBCO [8]. It is clear from this list that even if LZO buffers with excellent properties are actually produced by MOD [24] in a continuous line, many important questions remain unsolved in relation to the MOD process applied to LZO films.

In this research we aimed to study in details the role of the atmosphere on the crystallisation of LZO films deposited on Ni- 5at% W substrates from propionate solutions by spin coating. We used a cycle with one thermal step in which the pyrolysis was not treated separately from the crystallisation. The goal was to increase the crystallised part of the film, and to improve its texture. During the course of this research, we have also studied the reaction synthesis of LZO powders in order to provide some information on the crystallisation mechanism of bulk LZO.

## 2. Experimental

Lanthanum(III) 2,4-pentanedionate (La(acac)$_3$·3H$_2$O) and zirconium(IV) 2,4-pentanedionate (Zr(acac)$_4$) were used as solute. After weighing, a stoichiometric mixture of

La(acac)$_3$ and Zr(acac)$_4$ was introduced into propionic acid. The solution was placed in an ultrasonic bath for 15 min, then stirred and heated at 60 °C for 30 min. Finally, a transparent yellow precursor solution was obtained [25]. The concentration of the precursor solution was 0.5 mol/L (with respect to La$^{3+}$).

The precursor solution was dried at 100 °C to remove the solvent, yielding a dry precursor powder which was named hereafter the "as-received" powder. As-received powders were heat-treated at different temperatures for 30 min under Ar-5% H$_2$ atmosphere with the same heating rate (850 °C/h), the flow of gas was fixed at 0.5 l/h by a flow-meter. The structure of black residual powders was checked by x-ray diffraction (XRD), TEM (Philips 300 CX) and their chemical composition was also analyzed. Chemical analysis was performed on small samples (10-100 mg) at the central analysis centre of CNRS (CNRS-Vernaison Lyon, France). Metallic elements were determined by ICP in solution (with a relative error of 2%), C and H were quantified by catharometric detection (relative error 0.3 %), O was determined by measuring CO$_2$ produced by reaction of the pyrolysed products on an active C.

Ni-5%W substrates (size: 5 mm × 5 mm) were cut from a long tape (Evico GmbH), cleaned by acetone for 10 min in an ultrasonic bath, and covered by the precursor solution using spin-coating (rotation speed: 2500 rpm, rotation time: 30 s, acceleration: 3000 rpm/min), then the samples were dried at 80 °C for several minutes in air. samples were heated at 950 °C for 30 min under Ar-5% H$_2$ flow with a fixed heating rate of 850 °C/h. The films' thicknesses were measured by profilometry and by fitting the IR reflectivity. The films' thicknesses were typically in the range of 50-100 nm. Besides Ni-5at%W substrates, LaAlO$_3$ (100) single crystals (LAO) were also used in particular cases.

The texture of all samples was characterized by x-ray diffraction using a four-circle diffractometer (Siefert MZ IV) in the Schulz configuration. The diffractometer worked at Cu K$_\alpha$ wavelength with collimating multilayer optic from Xenocs (Grenoble, France) enabling a low divergent beam (horizontal angular divergence was 0.06° and the vertical one was 0.1°) with an irradiated area covering 2x1 mm$^2$. A graphite analyzer is mounted before the detector to prevent parasitic photon collection. This device is very efficient for measuring texture or studying single crystals or epitaxial thin films. In particular it has a very high signal/noise ratio enabling to detect diffracted peaks with weak intensities in the range of a few counts per second (c/s). Due to these intrinsic characteristics the diffracted intensity on polycrystalline is divided by 5 to 10 compared to more conventional apparatus (like D5000 Siemens in Bragg-Brentano scheme). Therefore, the reader must not be surprised by the low intensities recorded during this research. The crystalline quality of the surface of LZO films was checked by electron back scattered diffraction (EBSD) and RHEED.

XPS has been used to study the residual C at the surface of the films. The apparatus was a Axis Ultra from Kraos analytical Ltd. (GB). C (1s) has been registered versus sputtering time in the binding energy range 272-292 eV. The sputtering was performed with Ar ions accelerated at 4 kV under 15 mA providing a sputtering rate in the range of 0.1-0.2 nm/min.

### 3. Results

*3.1 Synthesis of bulk LZO powder under a reducing atmosphere*

Part of the as-received powders was heat treated under Ar-5% $H_2$ flow at 500 °C, 700 °C, 900 °C and 1100 °C to study its chemical composition evolution by chemical analysis and its crystallisation by XRD and by TEM.

The results of chemical analysis are reported in Table 1. Starting from a pentadionate precursor, it is obvious that the as-received powder was closer to a propionate than to a pentadionate. It agrees with the transformation occurring during their dissolution into a solution as already mentioned [15, 25]. During the heat treatment, H is completely eliminated while substantial amounts of C and O remain. The remaining oxygen points to the formation of an oxide; i.e. $La_2Zr_2O_7$ (LZO), as is certified by the weight loss analysis corresponding at 1100 °C to within 2% to the nominal composition of this oxide [25] when C is not taken into account in the molar fraction of the residue at 1100°C. The overall residual C level is very high, amounting 12 wt% even after heating at 1100 °C. Such large amount of carbon would correspond to approximately 6.6 mole of C per one mole of LZO. It should be noticed that the proportion of C increased slightly with heating due to the removal of H and O species during the conversion of the as-received powder into LZO.

XRD was performed on the heat treated powders, Fig. 1a. At 700 °C the powder was mainly amorphous with a unique diffracted peak at the position of the (111) peak of the fluorite form of LZO (SG Fm-3m, according to ICSD28991 data file). At 900 °C most of the peaks of the fluorite phase were visible. At 1100 °C all the peaks were indexed within the pyrochlore unit cell (SG Fd-3m according to ICSD15165). It means that the crystallisation of LZO powder from propionate precursor occurs in the temperature range 700-900 °C, that agrees well with what was found in films on Ni-5at%W substrates (800-850 °C) [13, 14, 15]. The structure of the oxide undergo a transformation from the fluorite phase to the pryrochlore phase at higher T, i.e. between 900 °C and 1100 °C. The fluorite phase is disordered and metastable in this temperature range [5].

TEM observations on these heat-treated powders confirmed that the onset of crystallisation of the as-received powder was at 700 °C, fig.1b. Some grains were already crystallised, and the electron diffraction pattern from a selected area confirmed that these grains were LZO. Fig.1c shows a TEM image of several grains after heating at 950 °C. All these grains had well organised facets and atomic planes in many grains were visible, testifying an almost complete crystallization. The mean grain size remains small, i.e. <10 nm. At 1100 °C, fig.1d, the TEM image shows that the grains had grown a little compared to grains synthesized at 950 °C, reaching about 10 nm while sintering has already started leading to large agglomerates (several hundreds of nm).

*3.2 Deposition of LZO film*

In a preliminary study, LZO films were deposited on $LaAlO_3$ (LAO) single crystals and heat treated at 950 °C for 30 min under air or Ar-5% $H_2$ flow. The case of air treatment is presented first. Except the reflexions from the substrate, the only diffracted lines in a θ-2θ diagram ( Fig.2a ) were observed at 2θ = 33.20° and 69.8° (not shown). This is a characteristic of a strongly **c** axis oriented LZO film. The (111) and (200) pole figures (not shown) proves the epitaxial nature of the film. Its structure may be fluorite or pyrochlore, the later one being characterised by {331} planes that are absent in the fluorite structure. The pole figure analysis of the {331} reflections evidencing 8 reflections at the tilt angle of χ = 46.5° due to (331) and equivalents and 4 reflections at χ = 76.7° due to (133) and equivalents, fig. 2b, demonstrates that the structure of this film is pyrochlore. So the peak at 2θ = 33.20° was indexed as (400) relative to the pyrochlore phase of LZO.

The diffracted intensity of the (400) peak of the samples which were heat treated under air flow is every high; surprisingly, the peak intensity dropped by a factor 5 to 10 when the atmosphere was switched to a reducing atmosphere (Ar-5%$H_2$). It indicates a less efficient crystallisation under very low oxygen partial pressure. In the case of a reducing atmosphere, the oxygen in the atmosphere surrounding the sample was due to residual leakage of the furnace and was measured by a mini zirconia gauge (from SETNAG-Marseille France) installed directly above the sample. At 950 °C the oxygen partial pressure was in the range of $10^{-20}$ atm. Curiously, under Ar-5%$H_2$, the diffracted intensity was sensitive to the speed of the gas flown over the sample and the line shifted to higher angle (2θ = 33.38°) indicating a deformed structure of LZO. The intensity increased by a factor of 2 when the speed of the gas was increased from $1.2 \times 10^{-4}$ m/s to $2.2 \times 10^{-3}$ m/s. The speed of the gas was approximated by the flow (imposed by a flow meter) divided by the section of the tube. In order to know if this phenomenon was linked with the crystallisation of LZO or with an effect of the substrate, several experiments were performed on Ni-5at%W substrates which are used in CCs applications.

Figure 3 shows the effect of the gas forming speed on the diffracted intensity of six films of LZO deposited as single layer on Ni-5at% W substrates. The samples were heat treated under the same conditions except that the speed of gas was changed by modifying the set point of the flow meter. The samples were measured using same parameters of the diffractometer enabling to record quite reproducibly the diffracted intensities (on identical samples, the measured intensities varies less than 10 %). All these ensure an authentic comparison of the samples. The first observation concerns the diffracted intensities which were small, much smaller than the one obtained on LAO single crystals (several tenths). It may suggest the influence of an inhibiting factor on the growth. Secondly, the LZO films were strongly **c** axis textured but with random contributions evidenced by the (222) diffracted line when the speed of gas was lower than $2.2 \times 10^{-3}$ m/s. Thirdly, the intensity of the (400) peak increased with increasing the speed of the gas flow. The increase (5 times) is much larger than experimental errors. Fourthly, the (222) contribution became negligible if the speed of the gas was sufficiently large, i.e. above $1.1 \times 10^{-3}$ m/s. Finally, the intensity of (400) peak just increases slightly when the speed of the gas is higher than $3.4 \times 10^{-3}$ m/s.

The best sample was characterized by XRD pole figure and EBSD which probes the sample's surface crystallinity. Fig. 4a shows the XRD pole figure of (222) which is a typical cube-textured LZO film grown epitaxially on the Ni-5at%W substrate. The epitaxial relationship is $[100]_{Ni}//[220]_{LZO}$ and $[001]_{Ni}//[001]_{LZO}$ as is usually observed by others. The in-plane rotation of LZO unit cell with respect to that of Ni is to better match both unit cells. Fig. 4b shows the same pole reconstructed from EBSD data. It clearly demonstrates that the surface of the film had the same texture. The sharpness of the texture was as good as that of the metallic substrate (typically 6.5° for in-plane texture, and 7.4° for out-of-plane texture). It is useful to mention that the RHEED pattern ( Fig. 5 ) of this film also shows that the ultimate surface (1 nm) was textured. The spots had some tendency toward a ring-shaped elongation in connection with the angular misalignment between grains. The planar misalignment was estimated by RHEED in the same range as those measured by XRD. The surface was rough at a scale of a few nanometers from AFM. Obviously, the crystallinity of the best films was good and could serve as a template for the epitaxial growth of other oxide film.

The C content of the top layer of LZO/Ni5W films has been probed by XPS after sputtering with Ar ions at a low rate of 0.1-0.2 nm/min. Samples processed under large gas flow were compared to samples processed at low gas flow and two annealing temperatures were compared. The data are combined in Fig. 6 where the C at% (vertical axis) is estimated from the photoelectron peak of C 1s (at 283 eV, binding energy), and the horizontal axis is the sputtering time. The figure shows a rather large C concentration at the top layer of LZO (4-7 at %), even after a treatment at 950°C for 40 min. Increasing the annealing temperature from 800 °C to 950 °C has not a spectacular effect, while increasing the gas speed over the sample decrease the surface C content. Even more, it shows that the C distribution is not homogeneous when the gas flow is small which agrees with amorphous or not textured areas in the depth of the sample.

**4. Discussion**

Measurements of the chemical composition of bulk powders treated under Ar-5%H2 to reproduce the conditions used for films synthesis showed a large amount of residual C (up to 12 wt %), trapped in the powder. Films of LZO also showed residual C at their surface but in lower proportion; the content and distribution of C seems to be affected by the gas flow. It is believed that this phenomenon is due to the heat-treatment atmosphere. Besides of our case, the phenomenon has been found in other MOD films deposited by CSD process [17, 26].

It is believed that the residual carbon is due to the phenomenon of carbon deposition which is well known in catalysts. The carbon deposition is an important reason for loss of catalytic activity. In the case of the partial oxidation of methane to produce syngas [27], authors discussed the possibility of decomposition of $CO_2$ or $CH_4$ to form carbon. In our case, the decomposition of propionate precursors emitted carbon-containing gases, for example, $CO_2$ or $CH_4$, and it is possible that these carbon-containing gases would be decomposed to form solid carbon according to Ref. 27. Because the residual carbon results from the decomposition of carbon-containing gases, and the decomposition of propionate precursors is almost finished at 800 °C [25], this can explain the reasons why there was an amount of residual carbon in final powders and why the amount of residual carbon changed slightly after heating above 700 °C.

TEM observation showed that the residual carbon existed at boundaries of LZO grains and not in LZO grains. The impurity in grain boundaries influences effectively the crystal growth during deposition of oxide films by CSD [16-18]. Other group results indicate that the residual carbon would influence the development of LZO film's texture. Usually, the free energy variation for nucleation is inversely proportional to the square of undercooling; therefore in chemical solution deposition process the crystallization driving force is very large [9, 10]. Although heterogeneous nucleation occurs more easily than homogenous nucleation and starts at the substrate-oxide interface, the homogenous nucleation may become competitive if the front of the epitaxial part cannot migrate rapidly due to some inhibiting factor. Based on this analysis, it is possible that some randomly oriented grains grew on the top layer of LZO film though the Ni-5at%W substrate owns a good cube texture, and then the texture of the final LZO film is cube texture with additional random orientations. Similar result has been found by others [17, 26]. This group demonstrated that the bottom layer of $CeO_2$ film deposited by CSD owned a cube texture while the top layer of the $CeO_2$ film could be random due to the residual carbon. Hence, the deposition of residual C seems to be a common phenomenon to CSD process when performed under reducing forming gas. It is essential then to have an appropriate procedure for removing it in order to obtain a well crystalline oxide film.

The main reason for the presence of residual carbon is the decomposition of carbon-containing gases during the pyrolysis of the LZO precursor film, two routes can be imagined to eradicate its influence. One is to increase the oxygen partial pressure to oxidize the residual carbon, but it could result in oxidizing the Ni-5at%W substrate [28]; so only applying a controlled oxygen partial pressure would be useful. We have preferred to use another simple way by increasing the flow speed of Ar-5%$H_2$ to dissipate the carbon-containing gases as rapidly as it can. The efficiency of this procedure has been demonstrated by our results (Fig. 3). According to knowledge about hydrodynamics, a boundary layer of C-containing gases should always exist at the surface even for large applied gas speed. Then the effectiveness of increasing the flow speed of Ar-5%$H_2$ gas has a physical limitation which has also been proved by Fig. 3.

Based on the above-mentioned analysis, it is clear that the crystallisation and texture of LZO films are dictated by two contributions, the epitaxial growth at the substrate/film interface and the growth of the random part, if the treatment conditions are not optimized. The LZO film with a well crystalline surface has been obtained when the carbon-containing gases were dissipated as rapidly as possible.

**V. Conclusion**

LZO formation of bulk powders and of films by CSD process have been studied using propionates treated in a reducing atmosphere (Ar-5%$H_2$). Large amount of residual carbon was found in LZO powders formed in these conditions (10 wt %). It was left from decomposition of carbon-containing gases emitted during the pyrolysis the precursor. Films of LZO were deposited on Ni-5at%W RABITS. The volume fraction of the cube texture was found to be a function of the speed of the gas flown above samples. Large speed of gas promotes an efficient dissipation of the carbon-containing gases, and it was a good method to eliminate the negative effect of the residual carbon in the crystallisation of LZO films. In optimized deposition conditions, LZO films with good surface crystallinity could be obtained on Ni-5at%W substrates. The existence of residual carbon in oxide films is a common question to films deposited by CSD processes under reducing condition.

**Acknowledgements**

This work was supported by the French embassy in Beijing and the International Laboratory for the Applications of Superconductor and Magnetic Materials (LAS2M-CNRS-NIN). It was also supported by the National High Technology Research and Development Program of China (863 Program) under contracts Nos. 2006AA03Z204 and 2008AA03Z202, and French ANR-Madisup project. Drs L. Ortega and C. Jimenez (respectively at Néel Institut and LMGP) are acknowledged for many helpful discussions. Drs De Sousa Meneses and B. Rousseau at CEMTHI (Orléans-France) are acknowledged for thickness measurements by IR reflectometry.

**Table caption**

Table 1      Elemental analysis of "as-received" LZO powders and of the fired powders after treatments at increasing temperatures under Ar-5%$H_2$ flow, with a corresponding gas speed of $1.2 \times 10^{-4}$ m/s. Nominal weight fractions for propionate and pentadionate are indicated for comparison.

**Figures caption**

Fig. 1 Crystallisation of bulk powders of LZO heat treated at different temperatures under Ar - 5%$H_2$ gas flow: a) XRD; b) TEM of a powder heated at 700°C; c) id at 950°C: d) id at 1100°C. White scale: 10 nm.

Fig. 2 XRD of LZO films crystallised at 950°C on LAO under static air or Ar-5%$H_2$ flowing at different speeds over the samples: a) XRD at tilt angle $\chi = 0$; b) pole figure evidencing the pyrochlore structure of LZO with 12 {331} equivalent reflections expected. Three reflexions are visible at $\chi = 76.7°$, and seven at $\chi = 46.5°$. The two missing reflections are barely visible at $\chi = 46.5°$, $\phi= 252°$ and $\chi = 76.7°$, $\phi= 252°$ respectively.

Fig. 3 XRD of LZO films deposited on Ni-5at%W, treated at 950°C under Ar-5%$H_2$ gas flow at different speeds of gas.

Fig. 4 XRD and EBSD pole figure on the best sample processed at the highest speed of gas ($6.8 \times 10^{-2}$ m/s).

Fig. 5 RHEED image of the LZO film which was deposited under optimized conditions.

Fig.6 C1s by XPS versus sputtering time for samples treated under different Ar-5%$H_2$ gas flow and at two temperatures.

```
 1
 2
 3
 4
 5
 6
 7
 8
 9
10
11
12
13
14
15
16
17
18
19
20
21
22
23
24
25
26
27
28
29
30
31
32
33
34
35
36
37
38
39
40
41
42
43
44
45
46
47
48
49
50
51
52
53
54
55
56
57
58
59
60
61
62
63
64
65
```

**Figure**

Fig. 1 Yu et al

1a)

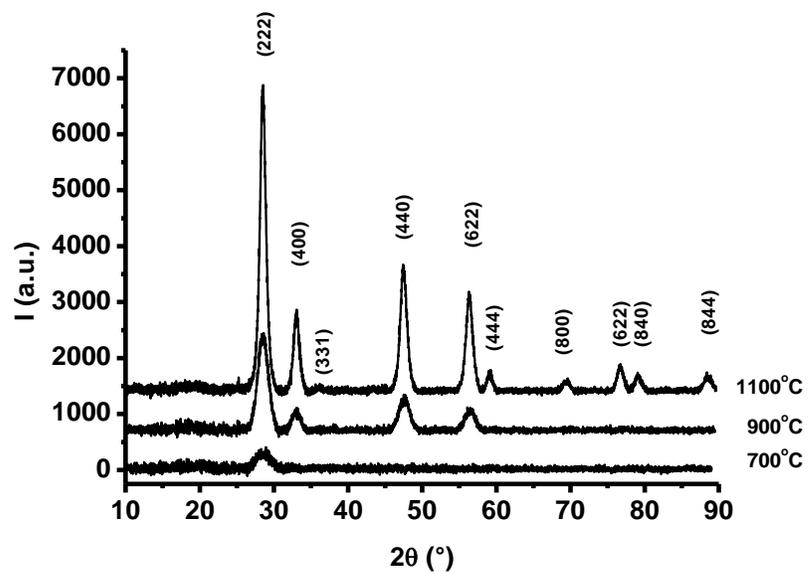

1b)

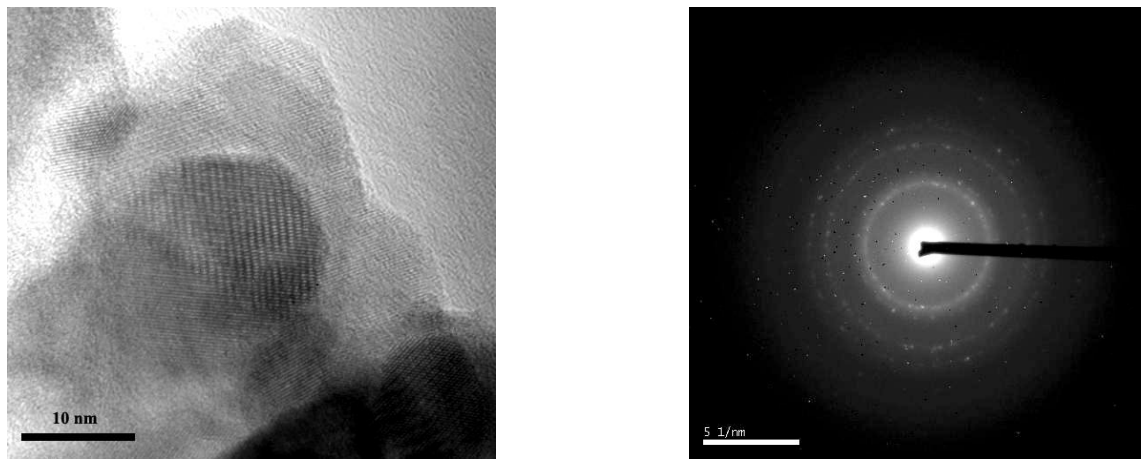

1c)

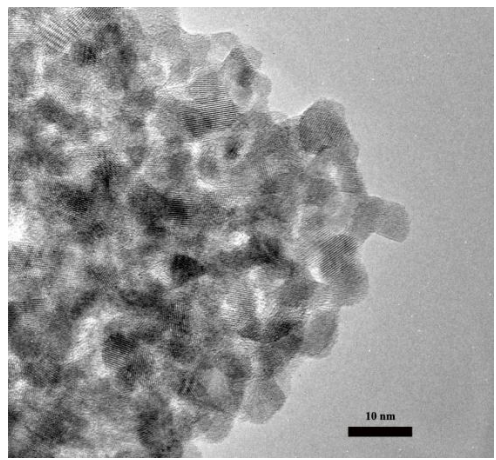

1d)

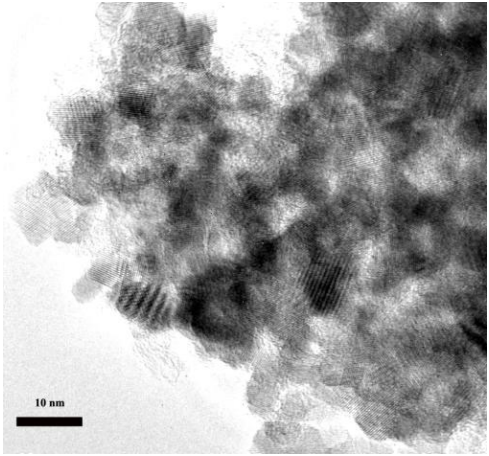

**Fig.2**

**(a)**

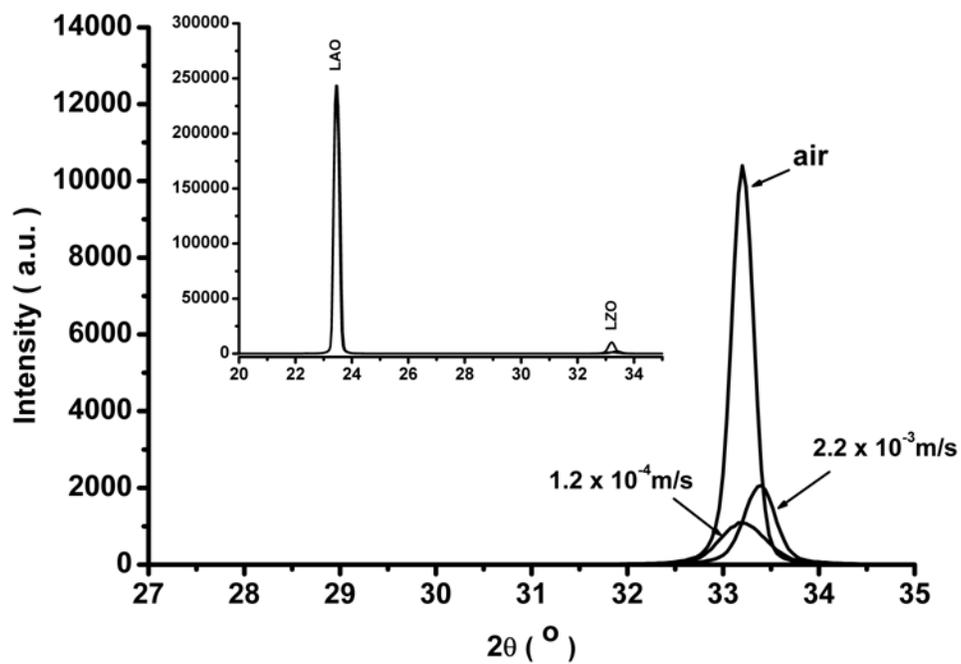

**(b)**

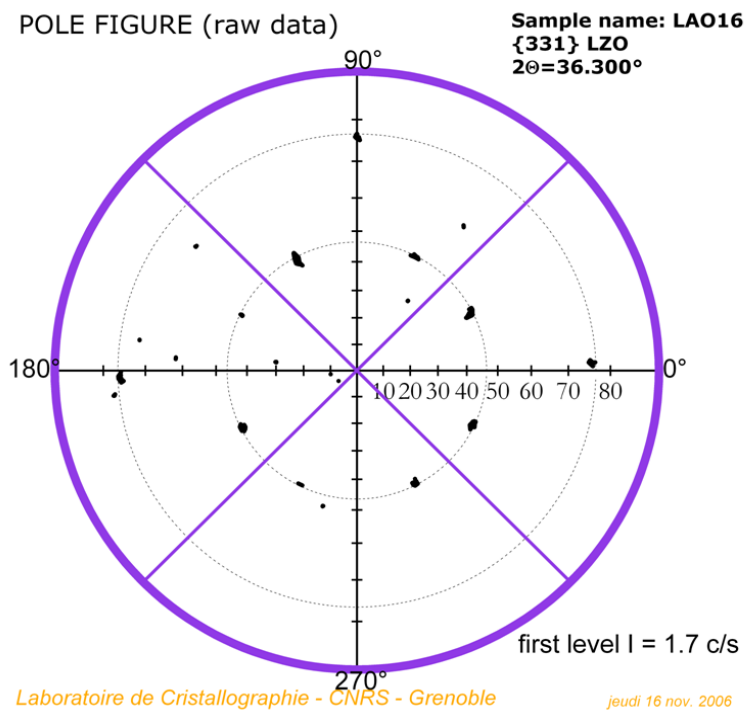

**Figure**

Fig.3

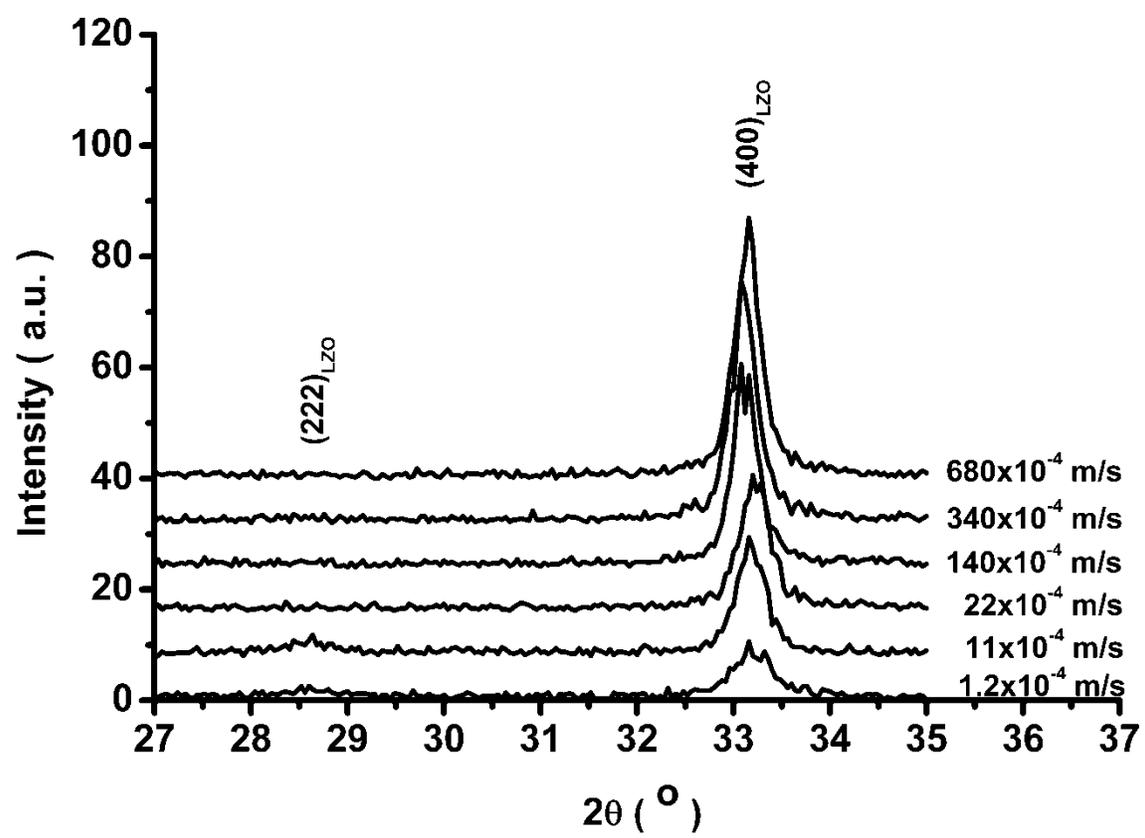

**Figure**

Fig.4 Yu et al

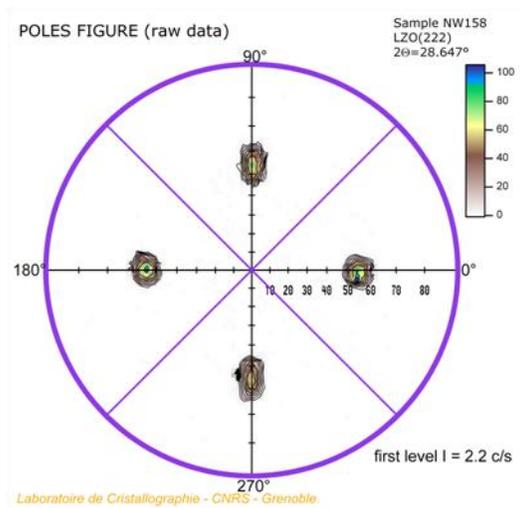 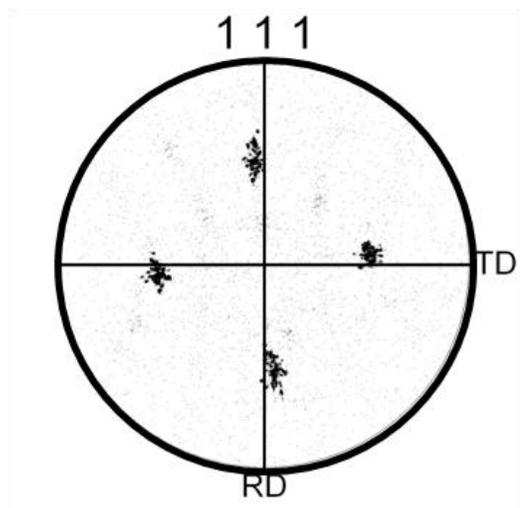

a) b)

Fig.5 Yu et al

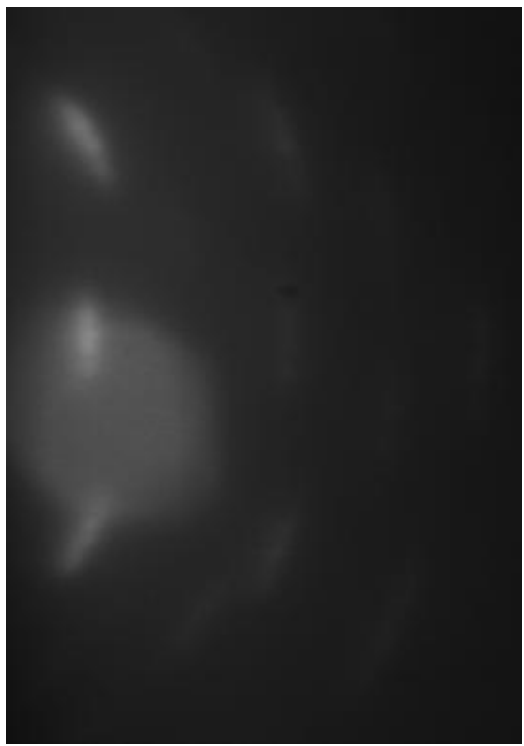

Fig.6

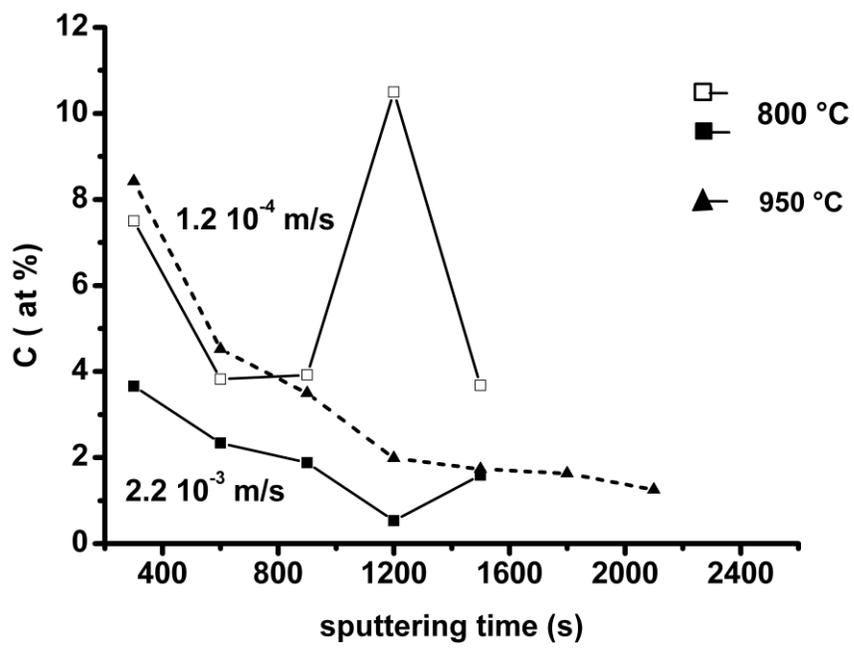

**Table**

Table 1 Yu et al

Table 1

| Temperature | Elements ( wt% ) | | | | |
|---|---|---|---|---|---|
| | C | H | O | La | Zr |
| *propionate* | (34) | (4.7) | (30.2) | (18.7) | (12.3) |
| *pentadionate* | (45.5) | (5.3) | (20.8) | (15) | (9.9) |
| **as received** | 33.4 | 4.8 | --------- | 18.63 | 11.39 |
| **500℃** | 9.67 | --------- | --------- | --------- | --------- |
| **700℃** | 10.7 | < 0.3 | 20.6 | 40.6 | 26.92 |
| **900℃** | 11.37 | --------- | --------- | --------- | --------- |
| **1100℃** | 11.99 | < 0.3 | 18.92 | 41.76 | 25.38 |

Note: (1) in brackets is the weight fractions calculated from the chemical formula; (2) --------- not determined.